\documentclass[aps,twocolumn,floats,prd,nofootinbib,nobibnotes,superscriptaddress,10pt]{revtex4-1}

\usepackage{graphicx,amsmath,amssymb,hyperref}
\hypersetup{colorlinks=true, linkcolor=red, citecolor=magenta, urlcolor=blue}
\usepackage{dcolumn}
\usepackage{multirow}
\usepackage{xcolor}
\usepackage{enumitem}
\usepackage{eqnarray}
\usepackage{makecell}
\usepackage{cancel}
\usepackage{etoolbox}

\newcommand{\lcdm}{\rm \Lambda CDM}

\newcommand{\mpci}{~{\rm Mpc}^{-1}}

\begin{document}
\title{Unveiling cosmological information on small scales with line intensity mapping}

\author{Sarah Libanore}%
\email{sarah.libanore@unipd.it}
\affiliation{Department of Physics, Ben-Gurion University of the Negev, Be'er Sheva 84105, Israel}
\affiliation{Dipartimento\,di\,Fisica\,e\,Astronomia\,G.\,Galilei,\,Universit\`{a} degli Studi di Padova,~Via Marzolo 8,\,35131 Padova,\,Italy}
\affiliation{INFN, Sezione di Padova, Via Marzolo 8, I$-$35131, Padova, Italy}

\author{Caner \"Unal}
\email{unal@post.bgu.ac.il}
 \affiliation{Department of Physics, Ben-Gurion University of the Negev, Be'er Sheva 84105, Israel}
\affiliation{Feza Gursey Institute, Bogazici University, Kandilli, 34684, Istanbul, Turkey}

\author{Debanjan Sarkar}
\email{debanjan@post.bgu.ac.il}
 \affiliation{Department of Physics, Ben-Gurion University of the Negev, Be'er Sheva 84105, Israel}

\author{Ely D. Kovetz}
\email{kovetz@bgu.ac.il}
 \affiliation{Department of Physics, Ben-Gurion University of the Negev, Be'er Sheva 84105, Israel}

\begin{abstract}
One of the toughest challenges in modern cosmology is to probe the small scales $k \gtrsim 0.5\mpci$ in the matter power spectrum and clustering. We show that such small scales will be accessible via upcoming line-intensity mapping surveys, with carbon monoxide (CO) emission from star-forming galaxies at high redshifts as an example. While these galaxies cannot be individually detected and the two-point correlations of the intensity fluctuation maps is not accessible at these scales, the voxel intensity distribution (VID) of the highly non-Gaussian intensity maps is sensitive to the integrated emission from faint sources. 
As we describe, the main limitations are due to uncertainties related with the halo mass function and the CO emission model. We show that via the VID, high-redshift next-generation experiments can probe deviations from $\lcdm$ of order unity, while stage-3 experiments will be able to probe deviations as small as $\lesssim10\%$ at least up to $k\sim 10 \mpci$.
\end{abstract}

\maketitle

\section{Introduction}
Integrated spectral-line emission from galaxies and the intergalactic medium probe the Universe up to the epoch of reionization (EoR). 
To measure it, numerous line-intensity mapping (LIM) surveys with different target lines are currently either proposed, under development or already online (see Refs.~\cite{Kovetz:2017agg,Bernal:2022jap} for review).
Maps built from their data will be highly non-Gaussian, due to the interplay between cosmology and astrophysics in determining the fluctuations in the sources' spatial distribution. Thus, information will have to be extracted from all point statistics, including (as first suggested in Ref.~\cite{Breysse:2015saa}) the voxel intensity distribution (VID), namely the one-point function of the observed voxels' intensities. 
Interestingly, the VID is sensitive to signal from both large and small scales, while higher order statistics (e.g.~the intensity power spectrum) are limited with respect to the latter because of the lack of angular resolution. 

The matter power spectrum (PS) on such small scales ($0.5\mpci \lesssim k \lesssim 10\mpci$) is not yet well constrained. Deviations from $\lcdm$ could arise e.g.\ from non-cold dark matter (DM) components~\cite{Murgia:2017lwo} or from features in the primordial power spectrum shape~\cite{Bridle:2003sa}. 
The number of probes investigating these scales is increasing and spans a range of observables, such as Milky Way satellites~\cite{Banik:2019smi}, cosmic shear~\cite{DES:2021wwk}, CMB lensing~\cite{CMB-S4:2016ple}, the UV luminosity function~\cite{Sabti:2021unj,Sabti:2021xvh}, CMB spectral distortions~\cite{Chluba:2012we,Nakama:2017xvq}, the stochastic gravitational-wave background~\cite{Inomata:2018epa,Unal:2020mts}, as well as LIM related studies, such as the Lyman-$\alpha$ forest~\cite{Chabanier:2019eai} and 21cm at cosmic dawn~\cite{Loeb:2003ya,Munoz:2019hjh}.
Another interesting LIM probe~\cite{Righi:2008br,Lidz:2011dx,Breysse:2014uia,Pullen:2012su,Breysse:2015saa,Breysse:2016szq} is emission from rotational carbon-monoxide CO($1\!\to\!0$) transition at $\nu_{\rm rest} \!= \!115.271\,$GHz. 
Sourced in star-forming molecular gas clouds, this line does not suffer from significant line-foreground contamination (see e.g.,~\cite{Chung:2017uot,COMAP:2021nrp}).
Currently, the Carbon Monoxide Mapping Array Project (COMAP) is releasing the Pathfinder data~\cite{COMAP:2021qdn}.~The observed frequency range of this detector is centered at $\nu_{\rm obs} \sim 30\,$GHz, making it optimized to study CO emission at $z \sim 3$~\cite{Li:2015gqa,COMAP:2021lae}. Several extensions have been proposed, aimed at decreasing the instrumental noise, increasing the observed redshift range or the field of view (see e.g., Refs.~\cite{Bernal:2019gfq,COMAP:2021nrp}), etc. 

In this work, we show that the VID of CO line-intensity maps can probe deviations from the linear matter PS in $\lcdm$ at least up to $k\sim 10 \mpci$.
The paper is organized as follows. Section~\ref{sec:formalism} describes how we parametrize the matter PS deviations and the models adopted for structure formation and CO emission. In section~\ref{sec:vid} we revise the VID formalism as firstly proposed in Ref.~\cite{Breysse:2015saa}, while in section~\ref{sec:analysis} we describe the statistical analysis set-up. To compute the VID we rely on the public release of the code\footnote{\url{https://github.com/jl-bernal/lim}} described in Ref.~\cite{Bernal:2019jdo}, that we suitably updated.  Section~\ref{sec:results} presents our results, including a first investigation on the VID constraining power on fuzzy DM. Finally, section~\ref{sec:conclusion} summarizes our conclusions.

\section{Formalism}\label{sec:formalism}

To describe model-agnostic deviations from $\lcdm$, we parameterize the linear matter PS as~\cite{Sabti:2021unj,Sabti:2021xvh}
\begin{equation}\label{eq:test}
    P(k) = \begin{cases}
    & a_1 P_{\lcdm}(k) \quad \text{for }\quad k < k_{\rm 12} \\
    & a_2 P_{\lcdm}(k) \quad \text{for }\quad k_{\rm 12}\leq k < k_{\rm 23} \\
    & a_3 P_{\lcdm}(k) \quad \text{for } \quad k_{\rm 23}\leq k < k_{\rm 34}\\    
    & a_4 P_{\lcdm}(k) \quad \text{for }\quad k \geq k_{\rm 34} \\
    \end{cases} \, .
\end{equation}
We set the fiducial values of $\{a_1,\,a_2,\,a_3,\,a_4\}$ to unity to recover structure formation in $\lcdm$. 

\noindent
The PS is related to the variance $\sigma^2(M_h,z)$ of the matter density field on different scales through
\begin{equation}\label{eq:sigmaM}
    \sigma^2(M_h,z) = \int_{k_{\rm min}}^{k_{\rm max}} \frac{\,dk}{2\pi^2}\, k^2P(k,z)W^2(M_h,\,k) \,,
\end{equation}
where, in our case, $[k_{\rm min},k_{\rm max}] \equiv [10^{-2},100]\mpci$.

The value $\sigma(M_h,z)\sim \delta_c \sim 1$ is used to define the range in which structure formation becomes non linear. Following this approach, in Eq.~\eqref{eq:test} we consider $k_{\rm 12}\!=\!0.5\mpci$, $k_{\rm 23}\!=\!2\mpci$ and $k_{\rm 34}\!=\!10\mpci$. 
When $k\! \lesssim \!k_{12}$, linear structure formation still holds nowadays and $\lcdm$ is well constrained by the {\it Planck 2018} CMB data analysis~\cite{Planck:2018nkj,Planck:2018vyg}. Therefore, we can safely assign a prior of $3\%$ to $a_1$~\cite{Chabanier:2019eai}.\footnote{Our analysis is stable with less informative priors on $a_1$, e.g., $10\%$. PS normalization uncertainties e.g.,~related with the $\sigma_8$ tension (see e.g.,~Ref.~\cite{Abdalla:2022yfr}) will not significantly affect the results.}
On the other hand, scales $k \!\gtrsim \!k_{\rm 34}$ are non-linear even at high redshift. Meanwhile, the ranges $k_{\rm 12}\! < \!k\! \leq \!k_{\rm 23}$ and $k_{\rm 23} \!<\! k\! \leq \!k_{\rm 34}$ are linear in the EoR and then become non-linear at $z\sim 3$. 
In Eq.~\eqref{eq:sigmaM} we consider the top-hat window function 
\begin{equation}\label{eq:window}
    W(k) = 3 \, \frac{\sin(kR) - kR\cos(kR)}{\left(kR\right)^3}\, ,
\end{equation}
on scale $R = \bigl(3M_h / 4\pi\rho_c(\Omega_m-\Omega_{\nu})\bigl)^{1/3}$, with $\rho_c$ the critical density for collapse and $\{\Omega_m,\,\Omega_\nu\}$ the matter and the neutrino density parameters. 
We then choose our fiducial halo mass function to be~\cite{Tinker:2008ff}

\begin{equation}\label{eq:dndm}
    \frac{dn}{dM_h}(z) = f(\sigma(M_h,z))\frac{\bar{\rho}}{M_h}\frac{d\log\sigma^{-1}(M_h,z)}{dM_h}\,,
\end{equation}
where $f(\sigma(M_h,z)) = A^T\bigl[1+(\sigma/b^T)^{-a^T}\bigr]\exp(-c^T/\sigma^2)$  and $\bar{\rho} \sim \rho_c\Omega_{m}$. The parameter $A^T$ normalizes the overall amplitude of the mass function, while $a^T,\,b^T$ define the tilt and amplitude of the low\,-\,mass power law and $c^T$ determines the high mass cutoff scale. We set the fiducial values of the parameters at $z=0$ and their redshift evolution based on Ref.~\cite{Tinker:2008ff}. The parameters $\{A^T,b^T,c^T\}$ are fixed throughout our analysis, while we marginalize over $a^T$ (with a $20\%$ prior, see the conclusions of Ref.~\cite{Tinker:2008ff}), to partially account for  mass-dependent deviations in the halo model. In principle, as the shape of the halo mass function, particularly at small masses, is still uncertain (see, e.g.,~\cite{McClintock:2018uyf}), accounting for this uncertainty is crucial. However, this is mitigated in our analysis because deviations from the fiducial choice in Eq.~\eqref{eq:dndm} would be reabsorbed by the CO luminosity function defined below. 

\begin{figure*}[ht!]
    \centering
    \includegraphics[width=2\columnwidth,height=0.85\columnwidth]{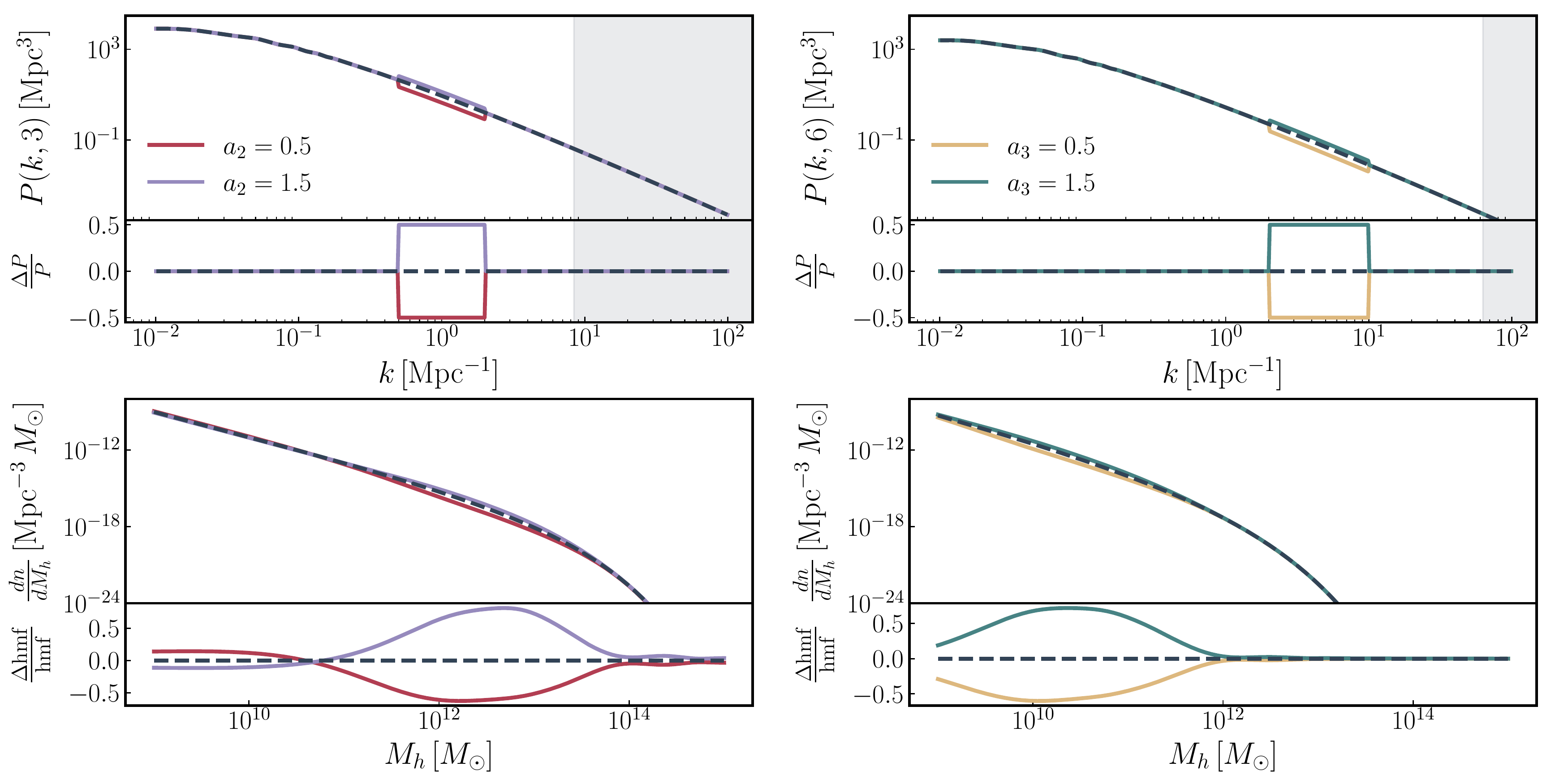}
    \caption{Example of how linear matter PS features, Eq.~\eqref{eq:test} (top row), propagate to the halo mass function, Eq.~\eqref{eq:dndm} (bottom row), in the ranges $z\sim 3$ (left) and $z\sim 6$ (right). Dashed lines indicate $\lcdm$ while shaded areas show where $\sigma(M_h,\,z)\sim 1$.}
    \label{fig:1}
\end{figure*}

Fig.~\ref{fig:1} compares $\lcdm$ linear matter PS and $P(k,z)$ in Eq.~\eqref{eq:test} with either decreased or increased power on small scales.
The figure also shows how perturbations reflect onto the DM halo mass function: at each $z$ in matter dominated era, each $k$ corresponds to a certain halo mass $M_h$ at the threshold to collapse (e.g.,~Refs.~\cite{Press:1973iz,Bond:1990iw,Sheth:1999mn,Tinker:2010my,Tinker:2008ff,Mead:2016zqy,Smith:2002dz}). 

\subsection{CO emission}\label{sec:CO_emission}
The DM halo mass distribution determines the star formation rate ($\rm SFR$), from which CO emission is usually modelled through semi-analytical prescriptions calibrated 
on simulations or data (Refs.~\cite{Bernal:2022jap,COMAP:2021nrp} summarizes the most popular models). 
We rely on the model originally introduced in Ref.~\cite{Li:2015gqa}, in which CO luminosity is related to infrared (IR) luminosity
as $L_{\rm CO}(M_h) \propto L'_{\rm CO}(M_h)\,(\nu_{\rm obs}/\nu_{\rm rest})^3$ $[{\rm K\,km\,s^{-1}pc^{-2}}]$, where
\begin{equation}
    L'_{\rm CO}(M_h) = \bigl[10^{-\beta} \,L_{\rm IR}(M_h)\bigr]^{1/\alpha} \,,
\end{equation}
and $L_{\rm IR}(M_h) = {\rm SFR}(M_h)/\bigl(10^{-10}\,\delta_{\rm MF}\bigr)$. 
The luminosity function of the CO emission is then computed as 
\begin{equation}\label{eq:dndl}
\begin{aligned}
&   \frac{dn}{dL_{\rm CO}}(L) = \int_{M_h^{\rm min}}^{M_h^{\rm max}}dM_h\,\frac{ e^{-L_{\rm cut}/L}}{L\sqrt{2\pi}\sigma_{\rm TOT}}\, \cdot
    \frac{dn}{dM_h}\, \cdot\\
 & \cdot \exp\biggl[-\frac{\bigl(\log L - \log(L_{CO}(M_h))+\sigma_{\rm TOT}/2\bigr)^2}{2\,\sigma_{\rm TOT}}\biggr]\,,
\end{aligned}
\end{equation}
where $\sigma_{\rm TOT}^2 = \sigma_{S}^2+\sigma_{\rm SFR}^2/\alpha^2$ and the parameters $\{\sigma_{\rm SFR},\,\sigma_S\}$ take into account the scatter both in the ${\rm SFR}-M_h$ and in the $L_{\rm CO}-{\rm SFR}$ relations. $L_{\rm cut}$ characterizes the exponential cut-off on low luminosities, which reflects the cut-off scale at the low mass end of $dn/dM_h$. 

In the analysis below, we consider as fiducial values for the astrophysical parameters the values described in Ref.~\cite{Li:2015gqa}, while we rely on Ref.~\cite{Behroozi:2012iw} for the ${\rm SFR}-M_h$ relation. 
This is degenerate with the ${\rm SFR}-L_{\rm IR}$ relation: intuitively, both higher ${\rm SFR}$ or larger IR emission would lead to an increased CO signal. Therefore, variations in $\delta_{\rm MF}$ can always be re-absorbed re-defining $\beta$; since our goal is not to constrain these parameters, we can safely fix $\delta_{\rm MF}$ to its fiducial value, provided that our observable (i.e.,~the VID in Sec.~\ref{sec:vid}) is affected by variations in both ${\rm SFR}-M_h$ and ${\rm SFR}-L_{\rm IR}$ relations through the parameter $\beta$. We note that $\beta$ is partially degenerate with $\alpha$ as well; we include both of them in the analysis since $\alpha$ directly relates with the CO emission we are interested in. 

To test how uncertainties in the astrophysics affect our results, in Sec.~\ref{sec:analysis} we also analyse the cases $\beta =\{-2.16,\,0.95\}$; we choose these values based on results in Ref~\cite{Li:2015gqa}. As we discussed, variations in $\beta$ can always be re-defined in $\{\alpha,\,\delta_{\rm MF}\}$ so to capture variations in all the ${\rm SFR}-M_h$, ${\rm SFR}-L_{\rm IR}$, $L_{\rm IR}-L_{\rm CO}$ relations. A more detailed investigation on the effect of different astrophysical models is left for a future, dedicated work.

\section{Voxel Intensity Distribution}\label{sec:vid}

It is evident from Eq.~\eqref{eq:dndl} that any variations induced by $P(k,z)$ to $dn/dM_h$ propagate also to $dn/dL_{\rm CO}$ and to the observed CO signal. This can be measured in each 3D voxel through the brightness temperature $T\propto \int dL\,[ L\,dn/dL_{CO}]$~\cite{Breysse:2016szq} and estimators have been built to measure its correlation functions (see Refs.~\cite{Bernal:2022jap,Schaan:2021hhy}). 

In particular, the one-point function, i.e.~the VID~\cite{Breysse:2015saa,Bernal:2022jap}, represents the histogram $B_i$ of the observed $T$ in bins $i$
\begin{equation}\label{eq:vid}
    B_i = N_{\rm vox}\int_{T_i^{\rm min}}^{T_i^{\rm max}}dT\,\mathcal{P}_{\rm TOT}(T)\,,
\end{equation}
where $N_{\rm vox}\!= \!\Omega_{\rm field}/\theta_{\rm FWHM}^2$ is the number of voxels in a given frequency channel,  $\theta_{\rm FWHM}$ is the angular resolution and $\Omega_{\rm field}$ the field of view. $\mathcal{P}_{\rm TOT}(T)$ is the probability distribution of observing the overall temperature $T$ in a single voxel, obtained by convolving the signal $\mathcal{P}(T)$ with the instrumental noise $\mathcal{P}_{n}(T)$. 

To compute the signal distribution in the map, we follow Ref.~\cite{Breysse:2016szq} and we consider $\mathcal{P}(T) = \sum_{N_{s}} p(N_{s})\mathcal{P}_{N_s}(T)$, where $p(N_{s})$ is the probability of having $N_s$ sources inside a voxel, while $\mathcal{P}_{N_s}(T)$ is the probability that, overall, they emit the observed temperature $T$. 
As Ref.~\cite{Breysse:2016szq} describes in details, the former combines a Poissonian and a lognormal distribution to describe the galaxy density field in terms of the clustering properties of the underlying matter field, while $\mathcal{P}_{N_s}(T)$ is computed recursively from  $\mathcal{P}_1(T)\propto dn/dL_{\rm CO}$, i.e.~the probability that in the voxel there is only one emitter with temperature $T$, which in turns directly depends on the CO luminosity function.

On the other hand, the instrumental noise, as discussed in Refs.~\cite{Breysse:2015saa,Bernal:2019jdo,Bernal:2019gfq}, can be modelled as a Gaussian
\begin{equation}
    \mathcal{P}_{n}(T) = \frac{1}{\sqrt{2\pi}\sigma_N}\exp\biggl[-\frac{T^2}{2\,\sigma_N^2}\biggr] \,,
\end{equation}
where the noise variance $\sigma^2_N$ is  given below in Eq.~\eqref{eq:sigmaN} and it depends on the properties of the survey considered. 
The primary effect of the instrumental noise on the VID is to wash out the information related with low $T$. Moreover, the system temperature $T_{\rm sys}$ determines the ground level for all the measurements, therefore decreasing the instrumental noise makes it possible to access lower $T$. 

\begin{figure*}[ht!]
    \centering
    \includegraphics[width=2\columnwidth,height=0.825\columnwidth]{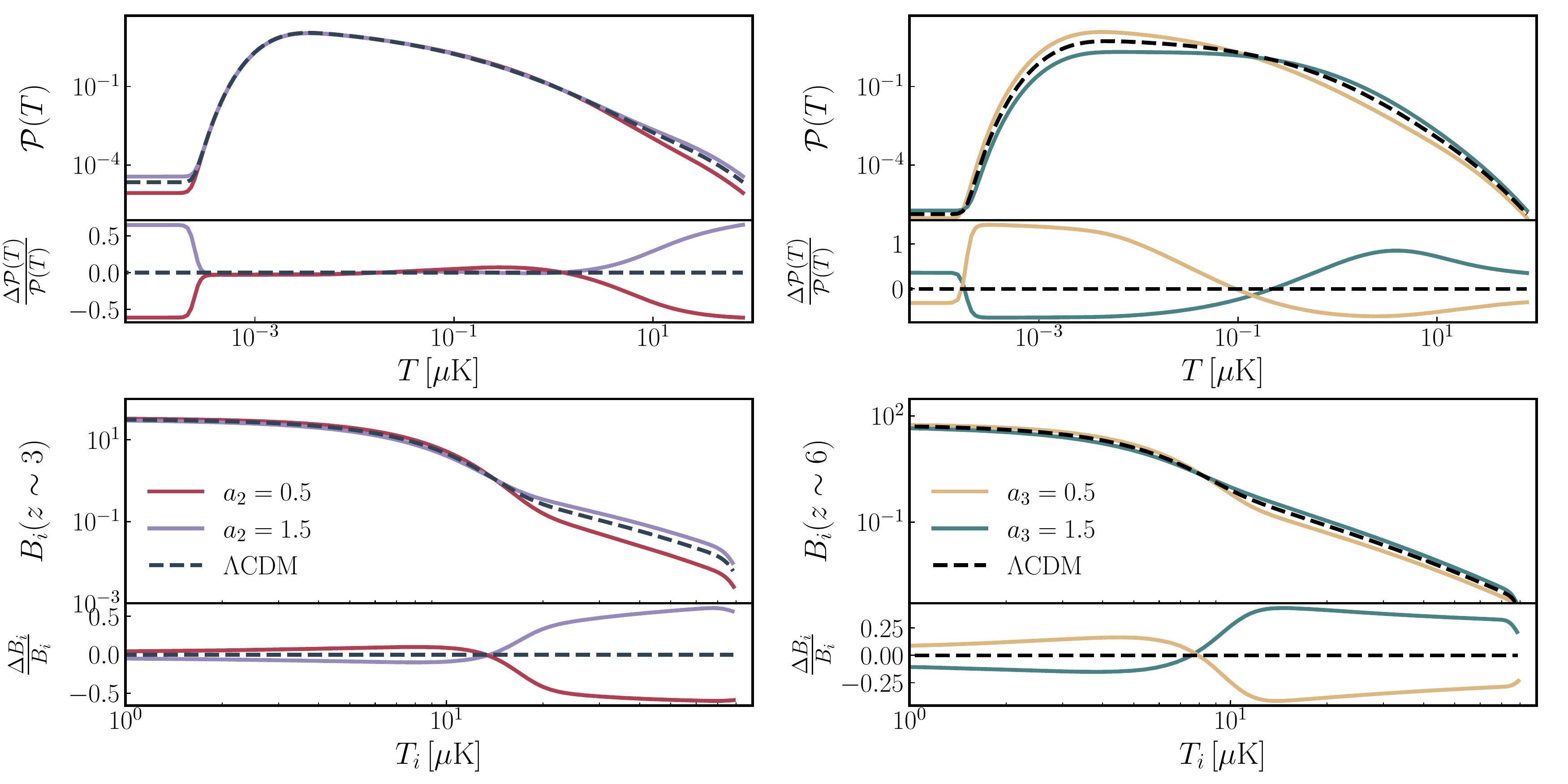}
    \caption{$\mathcal{P}(T)$ (top row) and VID $B_i$ from Eq.~\eqref{eq:vid} (bottom row) in our target redshift ranges $z\sim 3$ (left) and $z\sim 6$ (right). 
    The instrumental noise is determined by the parameters of the COS3 experiments in Table~\ref{tab:surveys}. The legend is the same as in Fig.~\ref{fig:1}.}
    \label{fig:2}
\end{figure*}
Finally, to estimate the VID from Eq.~\eqref{eq:vid}, it is important to properly choose the number of $T_i$ bins so as to maximize the information content that can be extracted from the analysis. We define $N_i = 200$ different temperature bins and we require that in each of them $B_i\geq 0.1$. Under this assumption, by stacking  $N_z$ redshift shells, in each frequency band our survey presents at least 100 voxels in each bin. 
Since noise-dominated bins are useless, we exclude the ones having $T_i < T^{\rm VID}_{\rm min} = 1\,\mu$K.

\section{Analysis set-up}\label{sec:analysis}

We compute forecasts for the planned COMAP-EoR survey~\cite{COMAP:2021nrp}, with a pair of instruments observing at redshift $z\!\sim\!3$ and three instruments observing at $z\!\sim\!6$. We also consider a deeper high-redshift survey (Deep), for which the observing time per pixel is increased to maximize sensitivity (analogously to the strategy used in CMB B-mode polarization observations~\cite{Jaffe:2000yt,Kamionkowski:2015yta}). For ultimate constraints, we consider a stage-3 experiment (COS3) with better sensitivity in both redshift ranges, due to a larger number of feeds (in analogy with~\cite{Bernal:2019jdo,Bernal:2019gfq}).

Table~\ref{tab:surveys} summarizes the main specifics of the different surveys, i.e.~the observed frequency range $\Delta\nu$, the field of view  $\Omega_{\rm field}$, the angular resolution $\theta_{\rm FWHM}$, the number of feeds $N_{\rm feeds}$ in each detector and the observational time $t_{\rm obs}$. For all  detectors we assume a $\delta\nu =2\,$MHz frequency resolution (we do not account for line-broadening in this work); this defines the number $N_z$ of redshift shells that can be observed with the tomographic CO survey. For more details on COMAP-EoR, see Ref.~\cite{COMAP:2021nrp}.

Table~\ref{tab:surveys} also reports the instrumental noise 
\begin{equation}\label{eq:sigmaN}
    \sigma_N = T_{\rm sys}\,\sqrt{\frac{N_{\rm vox}}{N_{\rm feeds}\, t_{\rm obs}\,\delta\nu}} \,,
\end{equation}
computed for each detector, where $ T_{\rm sys}$ is the system temperature. Note that with a fixed $t_{\rm obs}$, a large field of view $\Omega_{\rm field}$ increases $\sigma_N$, leading to a wider spread in the noise around its ground level $T_{\rm sys}$, i.e.~allowing for higher noise levels to occur. The vice-versa holds for a small field survey, such as the Deep COMAP-EoR version we entertain, in which the instrumental noise reaches lower values.

\begin{table}[h!]
    \centering
\renewcommand{\arraystretch}{1.3}
    \begin{tabular}{|c|*{6}{c}|}
    \hline
    & $\Delta\nu$ & $\Omega_{\rm field}$ & 
    $\theta_{\rm FWHM}$ &$N_{\rm feeds} $ & $t_{\rm obs}$   & $\sigma_N$ \\
    & [GHz] & [${\rm deg}^2$] &
    [arcmin] & \#& [hrs] & [$\mu$K] \\
    \hline
\multirow{2}{*}{\thead{ \smallskip EoR / COS3  \\ low-$z$}} & 
    30\,-\,34 & 4 & 4.5' & 19\,/\,$10^3$ &
    $5\,000$ &  $35$\,/\,5.4 
    \\
     & 26\,-\,30 & 4 & 3.9' & 19\,/\,$10^3$ &
    $5\,000$ &  $40$\,/\,4.7 \\    
         \hline
    \multirow{3}{*}{\thead{ \smallskip EoR (Deep)   \\  high-$z$}}
    & 17\,-\,20 & 4 (0.5) & 3.3' &  38 & 
    $7\,000$ & 22 (7.4)   \\
     & 15\,-\,17 & 4 (0.5) & 3.7' & 38 &
    $7\,000$ & 16 (5.5) \\
     & 13\,-\,15& 4 (0.5)& 4' & 38 &
    $7\,000\,$ & 14 (4.6)
    \\
    \hline
    \multirow{3}{*}{\thead{ \smallskip COS3 \\ high-$z$}}
    & 17\,-\,20 &  4 & 3.3' & $10^3$ &
    $7\,000$ & 4.3\\
     & 15\,-\,17 & 4 & 3.7' & $10^3$  &
    $7\,000$ &   3.1 \\
     & 13\,-\,15 & 4 & 4' & $10^3$  & 
    $7\,000$ &  2.7 \\
    \hline
    \end{tabular}
    \caption{Main detector specifics: 
    frequency range $\Delta\nu$, field of view $\Omega_{\rm field}$, angular resolution $\theta_{\rm FWHM}$, number of feeds, 
    observational time $t_{\rm obs}$. More data on COMAP-EoR in Ref.~\cite{COMAP:2021nrp}. Instrumental noises $\sigma_N$ are computed through Eq.~\eqref{eq:sigmaN}.}
    \label{tab:surveys}
\end{table}

Fig~\ref{fig:2} shows $\mathcal{P}(T)$ and $B_i$ for $\lcdm$ and the model in Eq.~\eqref{eq:test} (the $\{a_i\}$ parameters have the same values as in Fig~\ref{fig:1}), as observed by the COS3 detectors. Note that, since $\mathcal{P}(T)$ is a normalized probability distribution, any dips and hills on certain scales in the linear matter PS affect the distribution of sources at all $T$.

\subsection{Fisher matrix}
To compute forecasts for the CO surveys previously described, we use the Fisher matrix for the VID~\cite{Breysse:2015saa}
\begin{equation}\label{eq:fisher_vid}
    F_{\alpha\beta} = \sum_{D} \sum_{i > T_i^{\rm min}} \frac{1}{\sigma_i^2}\frac{d(N_z B_i)}{d\theta_{\alpha}}\frac{d(N_zB_i)}{d\theta_{\beta}} \,,
\end{equation}
where $D$ is the number of detectors allowed in the survey (e.g.~$D = 2 (3)$ for low-$z$(high-$z$) detectors, respectively) and $i$ defines the temperature bins in which $B_i$ is computed. In Eq.~\eqref{eq:fisher_vid}, we  stack all the $N_z$ observed redshift shells before performing the analysis, so to consider their cumulative signal. Note that, at this level, we are assuming that in each of the $D$ detectors of each survey the astrophysical evolution of the signal is negligible and all the $\delta\nu$ frequency channels (i.e.~all the redshift shells in a certain frequency band) have the same expected VID. 

\noindent
The expected variance of the signal in Eq.~\eqref{eq:fisher_vid} is computed  assuming a Poisson distribution~\cite{Breysse:2015saa}, in which $\sigma_i^2\! =\! N_zB_i$. 
As Ref.~\cite{Sato-Polito:2022fkd} demonstrates, the contribution of the cosmic variance is subdominant with respect to $\sigma_i^2$ for our ranges of system temperature and angular resolution.

The parameter set $\theta$ in Eq.~\eqref{eq:fisher_vid} contains, besides $\{a_i\}$ from Eq.~\eqref{eq:test}, the halo mass function parameter from Eq.~\eqref{eq:dndm} and the astrophysical parameters that model the CO signal: $\{a^T,\,\alpha,\,\beta,\,\sigma_{\rm SFR},\,\sigma_{S},\,L_{\rm cut}\}$. We use fiducial values $\{1.47\,(1+z)^{-0.06}, 1.37, -1.74, 0.3, 0.3, 50\,L_\odot\}$. 
The cosmological parameters are fixed to the {\it Planck 2018} fit~\cite{Planck:2018vyg}, as their effect is either subdominant or captured by other parameters. For example, variations in $a^T$ also account for $\mathcal{O}(5\%)$ deviations in $\Omega_m$ that propagate to the halo mass function and change $B_i$ by $\mathcal{O}(\lesssim 40\%)$. 

\section{Results}\label{sec:results}

\begin{figure}[ht!]
    \centering
   \includegraphics[width=\columnwidth]{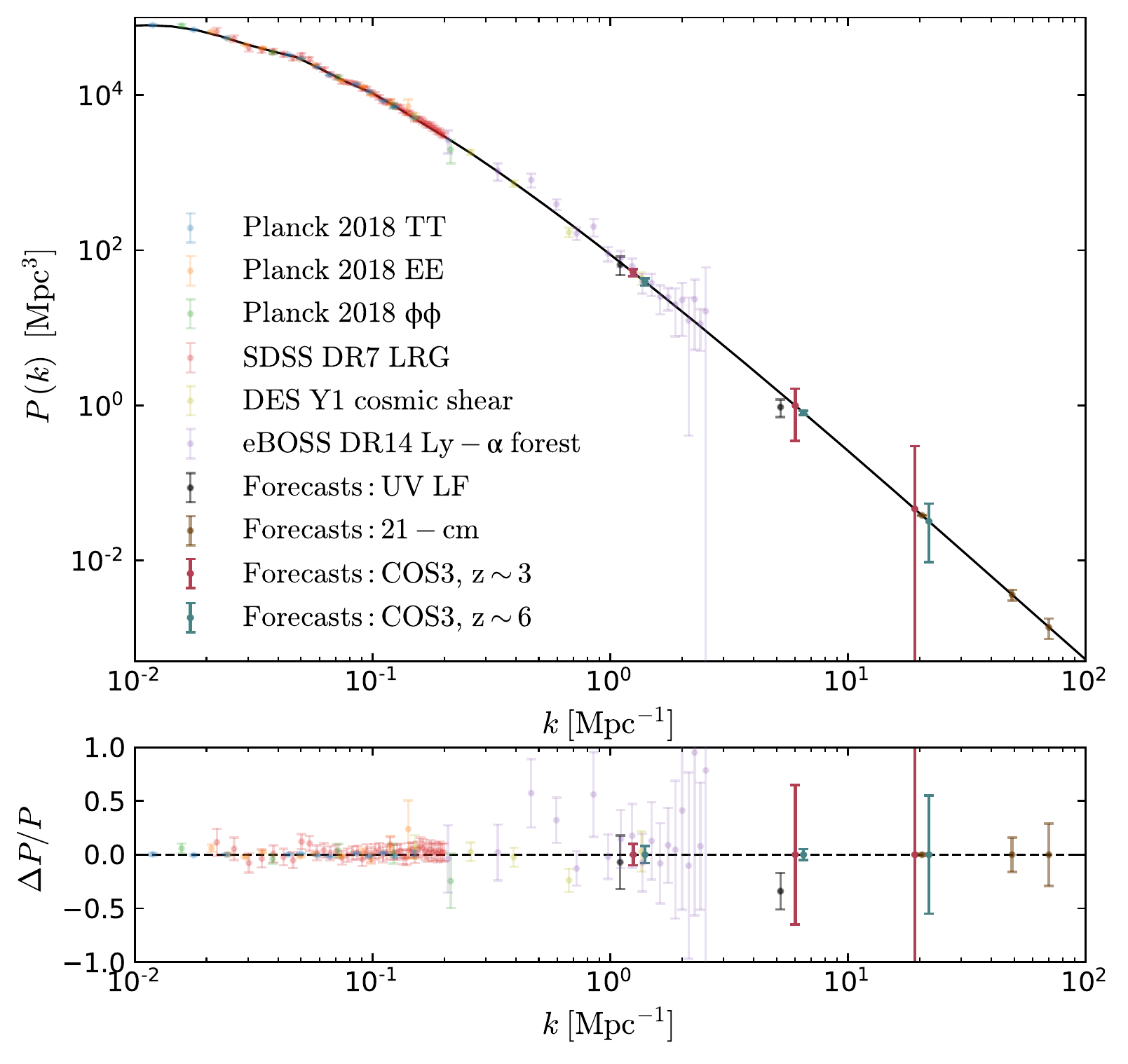}
    \caption{Our $1\!-\!\sigma$ VID constraints on $\{a_2,\,a_3,\,a_4\}$ linearly projected to $P(k,0)$ for the COS3 surveys in Table~\ref{tab:surveys}. $\{a_2,a_3\}$ error bars are plotted at bin centers, $a_4$ arbitrarily at $\sim 20\,{\rm Mpc}^{-1}$. Other constraints from Refs.~\cite{Chabanier:2019eai,Sabti:2021unj,Munoz:2019hjh}.}
    \label{fig:pk}
\end{figure}

Table~\ref{tab:constraints_pk} shows forecasts for the surveys in Table~\ref{tab:surveys} and Fig.~\ref{fig:pk} shows the linear matter PS with $1\!-\!\sigma$ marginalized errors for $\{a_2,\,a_3,\,a_4\}$ for the COS3 surveys.
\begin{table}[ht!]
\renewcommand{\arraystretch}{1.3}
    \centering 
    \begin{tabular}{|c|c|c|c|}
    \hline
    & {$a_2$} & {$a_3$} & {$a_4$}\\
\hline
    COS3 low-$z$ & 22\%  & 68\%  & 624\%  \\
    \hline
    EoR (Deep) high-$z$ & 107\% (32\%) & 152\% (19\%) & (345\%) \\
    \hline
    COS3 high-$z$ & 10\% & 6\% & 64\% \\
    \hline
    \end{tabular}
    \caption{$1\!-\!\sigma$ VID constraints on the $P(k,0)$ parameters for each of the surveys described in Table~\ref{tab:surveys}. We only show uncertainties $\leq\! 1000\%$ (and so omit COMAP-EoR low-$z$, although a Deep version of it would yield $\mathcal{O}(100\%)$ bounds on $a_{2}$). All results consider $3\%$ prior on $a_1$ and $20\%$ prior on $a^T$.}
    \label{tab:constraints_pk}
\end{table}

\begin{figure}[ht!]
    \centering
   \includegraphics[width=\columnwidth]{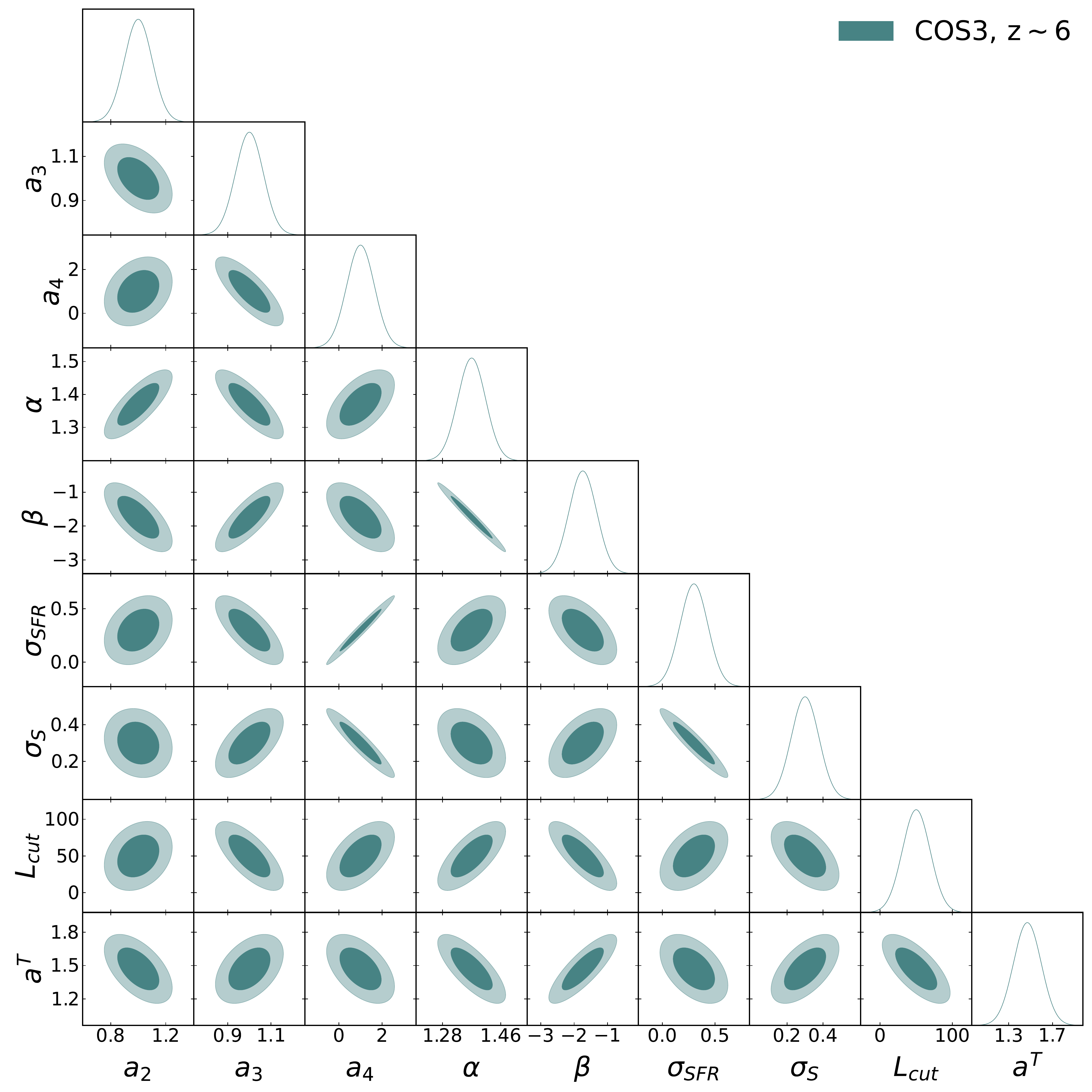}
    \caption{$1-$ and $2-\sigma$ VID constraints on $\{a_2,\,a_3,\,a_4\}$ and astrophysical parameters for COS3 at $z\sim6$.}
    \label{fig:ellipse}
\end{figure}

\noindent
To stress the importance of accessing the PS on a wide $k$ range,
we show state-of-the-art constraints from Ref.~\cite{Chabanier:2019eai} and forecasts from other probes on comparable scales.
We refer to Ref.~\cite{Sabti:2021unj} for the UV luminosity function and Ref.~\cite{Munoz:2019hjh} for 21-cm with moderate foregrounds. 
Our results indicate that in overlapping scales, LIM has the potential to compete with and even surpass the constraining power of other probes; also, COS3 high-z results are fairly competitive with 21-cm results from Ref.~\cite{Munoz:2019hjh} once we fix $\{k_{\rm 23},k_{\rm 34}\} $ in Eq.~\eqref{eq:test} to the same bin $k\in[3,38]\mpci$.
As Fig.~\ref{fig:ellipse} highlights, this reach is essentially limited by degeneracies with parameters characterizing the smallest sources that can emit CO, mostly the luminosity (mass) cut-off and the SFR scatter. 

\noindent
To check the stability of our results against astrophysical uncertainties, we analysed different fiducials for $\{\alpha,\,\beta\}$ (which are partially degenerate, see Fig.~\ref{fig:ellipse}): for $\{1.37,\,-2.16\}$ or $\{1.17,\,0.21\}$ (higher or lower signal, chosen according to~\cite{Li:2015gqa}), $\{a_2,\,a_3\}$ forecasts change $\lesssim 2\%$.

All of our results consider intensity maps that contain just signal and instrumental noise.
To be more conservative, we also consider the case where the brightest voxels in the map are masked (e.g.~to remove line-interloper foregrounds~\cite{Breysse:2015baa,Breysse:2016szq}). Masking the $3\%$ brightest pixels, our $\lesssim\!10\%$ constraints degrade to $\mathcal{O}(100\%)$, but we do not completely lose sensitivity, even at the $k \lesssim 10\mpci$ scales (where order unity constraints are still quite meaningful, in the absence of other information). We emphasize, though, that CO is not expected to have significant line-foreground contamination~\cite{Chung:2017uot}. 
Meanwhile, to partially account for continuum foregrounds, we subtract the mean from all maps~\cite{Bernal:2019jdo,Bernal:2019gfq}.

\subsection{Fuzzy Dark Matter}

To further stress the capability of the VID in constraining cosmological models, we present a first investigation related with fuzzy dark matter (FDM), namely the effect of ultra-light axions on structure formation (see e.g., Refs.~\cite{Preskill:1982cy,Dine:1982ah,Kawasaki:2013ae,Hlozek:2014lca,Marsh:2015xka,Hui:2016ltb,Desjacques:2017fmf}). Depending on the fraction $f_{\rm FDM}$ of DM of this kind, small scales in the matter PS get suppressed; the cut-off scale where this happens depends on the mass $m_{\rm FDM}$ of the particle considered. 

\begin{figure}
\includegraphics[width=\columnwidth]{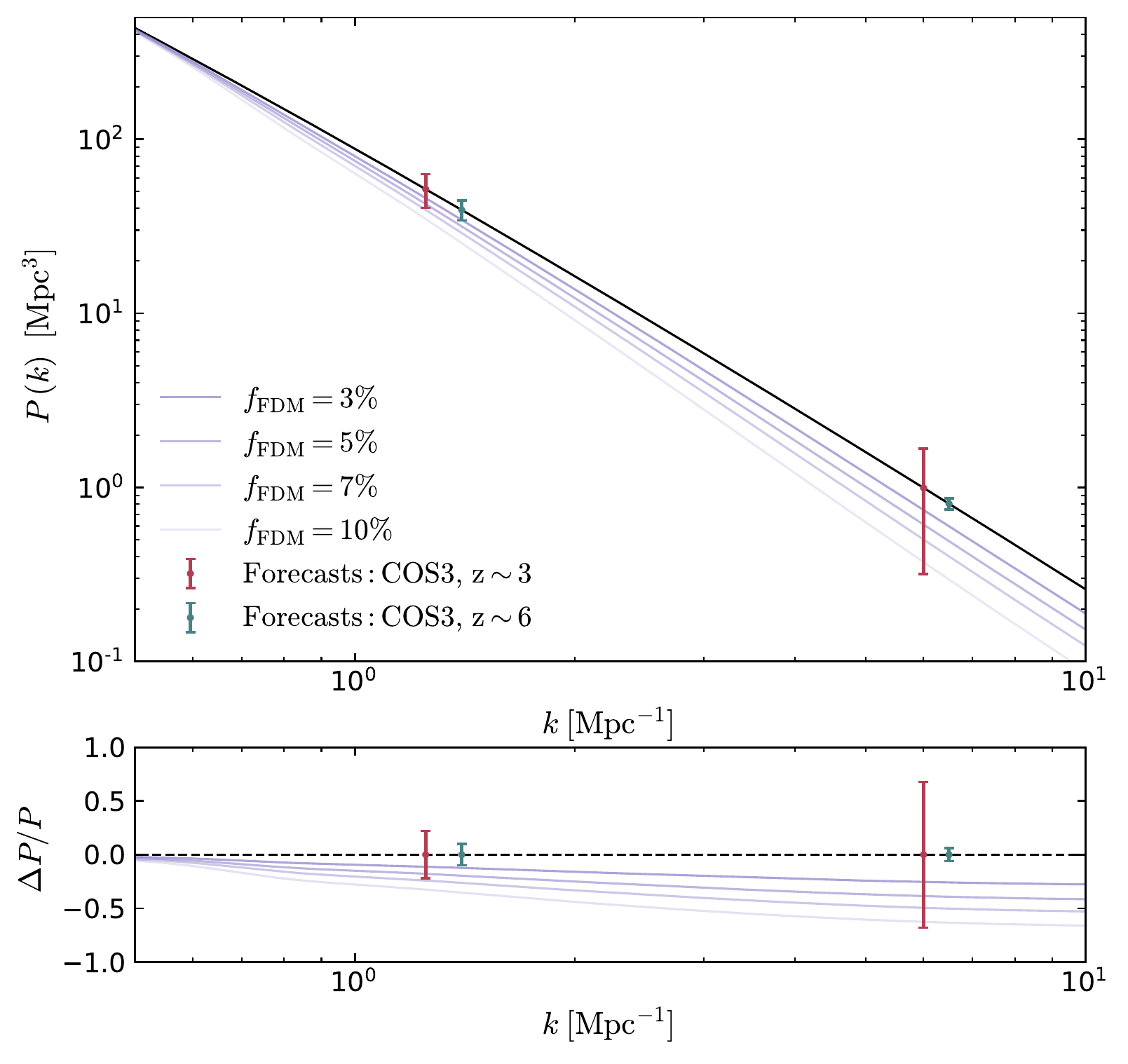}
\caption{$\lcdm$ matter power spectrum (black line) compared with matter power spectra when $m_{FDM}= 10^{-24}\,\rm eV$ and $f_{\rm FDM} = \{0.03, 0.05, 0.07,0.1\}$ (colored lines, computed through \texttt{AxionCAMB}\footnote{\url{https://github.com/dgrin1/axionCAMB}}~\cite{Hlozek:2014lca}) in the $k$-range probed by the parameters $a_2,\,a_3$ defined in Eq.~\eqref{eq:test}. The red and green errorbars show respectively our marginalized $1-\sigma$ errors on $\{a_2,\,a_3\}$, in agreement with the $\chi^2$ results described in the text.}
\label{fig:fdm}
\end{figure}

As figure~\ref{fig:fdm} shows, for $m= 10^{-24}\,\rm eV$, where current constraints still allow $f_{\rm FDM}$ to be large (see Refs.~\cite{Flitter:2022pzf,Unal:2022ooa} and references therein), the cut-off scale falls in the range probed by the $\{a_2,\,a_3\}$ parameters defined in Eq.~\eqref{eq:test}. 
Deviations induced by FDM in the matter PS can then be probed through the analysis described in this work. To provide a rough estimate of the constraining power of our approach, we compute the reduced $\chi^2$ as
\begin{equation}\label{eq:chi2}
\begin{aligned}
	\chi^2 =& \frac{1}{\rm DoF}\biggl[\sum_{k\in[k_{12},k_{23}[}\frac{\bigl(P_{\lcdm}(k)-P_{\rm FDM}(z)\bigr)^2}{\sigma_{s,a_2}^2} \\
	&+ \sum_{k\in[k_{23},k_{34}[}\frac{\bigl(P_{\lcdm}(k)-P_{\rm FDM}(z)\bigr)^2}{\sigma_{s,a_3}^2}\biggr]\,,
\end{aligned}
\end{equation}
where $P_{\lcdm}(k),\,P_{\rm FDM}(k)$ are the matter PS respectively in $\lcdm$ and in the presence of FDM, while $\sigma_{s,a_2},\,\sigma_{s,a_3}$ are the marginalized errors on $\{a_2,\,a_3\}$ for the surveys $s=$ COS3 low-$z$ and COS3 high-$z$ (see Table~\ref{tab:constraints_pk}). The number of degrees of freedom ($\rm DoF$) is computed as the sum of the number of detectors $D$ in the survey, the number of redshift shells $N_z$ and the number of independent $k$ modes observed in the intervals $k\in[k_{12},\,k_{23}[$ and $k\in[k_{23},\,k_{34}[$. Our $\chi^2$ results show that, if $m_{\rm FDM} = 10^{-24}\,\rm eV$, COS3 low-$z$ and high-$z$ can detect $f_{\rm FDM} \simeq 10\%$ and $\simeq 6\%$ at $2-\sigma$ level respectively. Our results are comparable with forecasts for future 21-cm surveys recently described in Ref.~\cite{Flitter:2022pzf}.

\section{Conclusions}\label{sec:conclusion}

In this work, we propose to use the voxel intensity distribution of line-intensity mapping surveys as a tool to constrain $\lcdm$ deviations in the matter power spectrum. We analyse the constraining power of planned and future surveys targeting CO both at low redshift ($z\sim 3$) and high redshift ($z\sim 6$). We find in particular that high-redshift next-generation experiments can probe deviations from $\lcdm$ of order unity (or better when the field of view is small enough), while stage-3 experiments will be able to probe deviations as small as $\lesssim10\%$ at least up to $k\sim 10 \mpci$. This will allow us to probe different cosmological models; as a first, illustrative example, we showed that the VID can constrain fuzzy dark matter with $m_{\rm FDM} = 10^{-24}\,\rm eV$ up to $\mathcal{O}(6\%)$.

Our analysis likely {\it underestimates} the sensitivity of the (CO) VID to the small scale power. 
We consider only three out of four detectors at high redshift (SFR estimates at the highest redshift bin are too uncertain). 
Moreover, we do not include the power spectrum of the maps in this analysis: while LIM surveys resolution is far from enough to probe clustering on 
small scales, the power spectrum will be useful in constraining 
the line-luminosity function 
(clustering and shot noise are sensitive to its first and second moments~\cite{Bernal:2022jap}), potentially breaking degeneracies in  
VID estimates and increasing the sensitivity to cosmological information~\cite{COMAP:2018kem,Sato-Polito:2022fkd}. 
We are also quite conservative in the choice of $M_h^{\rm min}$ and the SFR at high redshift (recent results from JWST may indicate that star formation and specifically line emission is significant already at very high redshift~\cite{2022arXiv220712657C,2022arXiv220712474F,2022arXiv220712430S,2022arXiv220712388T,2022arXiv220712356D,2022arXiv220712338A,2022arXiv220711379S,2022arXiv220711259S,2022arXiv220711217A,2022arXiv220713693K}). 

Potential uncertainties however comes from the choice of the astrophysical CO emission model and on the assumptions on the emitter distribution (e.g. their redshift evolution and bias). To overcome this issue, it is important to analyse how the overall CO signal gets affected by different choice of parameters and to understand whether using different VID formalism (e.g. the ones recently proposed in~\cite{Breysse:2022alx,Chung:2022zeu,Breysse:2022fdi}) can help breaking degeneracies between them; we leave this for a future, dedicated work. 

Finally, in our analysis we consider only CO, while many lines could be targeted, especially at high redshifts, including  fine structure lines ([CII], [OII],[OIII], [NII], etc.) and the hydrogen $H_\alpha$, $H_\beta$ and Lyman-$\alpha$ lines, among others~\cite{Bernal:2022jap}. Modeling multiple lines is more involved~\cite{Serra:2016jzs,Sun:2019multi,Mas-Ribas:2022jok,Sun:2022ucx}, but the potential to overcome degeneracies and model uncertainties (e.g.~by using line ratios~\cite{Solomon:2005xc,Kewley:2019ratio}) and mitigating cosmic variance via the multi-tracer 
approach~\cite{Seljak:2008xr,McDonald:2008sh,Kannan:2021ucy,Schaan:2021gzb} are quite promising.

To conclude, via the (CO) VID, we showed that high-redshift LIM will enable the study of $\lcdm$ on small scales. Future work will explore implications for non-Gaussianity~\cite{Unal:2018yaa,Sabti:2020ser}, spectral-index running~\cite{Munoz:2016owz}, DM models~\cite{Carlson:1992fn,Hu:2000ke,Bode:2000gq,Hlozek:2014lca,Flitter:2022pzf,Inman:2022uvy}, primordial magnetic fields~\cite{Planck:2015zrl}, baryon feedback~\cite{Bocquet:2015pva,Castro:2020yes}, other halo mass functions~\cite{Crocce:2009mg,Jenkins:2000bv,Warren:2005ey,Watson:2012mt} and CO emission~\cite{Bernal:2022jap,COMAP:2021nrp}, through novel VID prescriptions~\cite{Breysse:2022alx} and multi-line cross correlations~\cite{Chung:2022zeu,Breysse:2022fdi}.

\bigskip
\acknowledgments
\noindent The authors thank Jos\'e Luis Bernal for useful discussions and assistance with the {\tt lim} package.  We thank Jordan Flitter, Tal Adi, Jos\'e Luis Bernal and the anonymous referees for extremeley useful comments on the manuscript. SL~was supported by the Council for Higher Education of Israel's PhD Sandwich Scholarship Program and the Fondazione Ing.\ Aldo Gini scholarship. 
CU~is supported by the BGU Kreitman fellowship, the Excellence fellowship of the Israeli Academy of Sciences and Humanities, and the Council for Higher Education. EDK is supported by an Azrieli Foundation Faculty Fellowship.

\end{document}